\documentclass[aps,prl,twocolumn,superscriptaddress]{revtex4-2}
\usepackage{graphicx}
\usepackage{amsmath}
\usepackage{hyperref}
\usepackage{amssymb}     
\usepackage{color}       

\begin{document}


\title{First $^{94}$Nb($n,\gamma$) Measurement: Constraining the Nucleosynthetic Origin of $^{94}$Mo in Presolar Grains}

\author{J.~Balibrea-Correa} \affiliation{Instituto de F\'{\i}sica Corpuscular, CSIC - Universidad de Valencia, Spain} %
\author{J.~Lerendegui-Marco} \affiliation{Instituto de F\'{\i}sica Corpuscular, CSIC - Universidad de Valencia, Spain} %
\author{C.~Domingo-Pardo} \affiliation{Instituto de F\'{\i}sica Corpuscular, CSIC - Universidad de Valencia, Spain} %
\author{V.~Babiano-Suarez} \affiliation{Instituto de F\'{\i}sica Corpuscular, CSIC - Universidad de Valencia, Spain} %
\author{I.~Ladarescu} \affiliation{Instituto de F\'{\i}sica Corpuscular, CSIC - Universidad de Valencia, Spain} %
\author{M.~Krti\v{c}ka} \affiliation{Charles University, Prague, Czech Republic} %
\author{G.~Cescutti} \affiliation{Istituto Nazionale di Fisica Nucleare, Sezione di Trieste, Italy} \affiliation{Department of Physics, University of Trieste, Italy} %
\author{S.~Cristallo} \affiliation{Istituto Nazionale di Fisica Nucleare, Sezione di Perugia, Italy} \affiliation{Istituto Nazionale di Astrofisica - Osservatorio Astronomico d'Abruzzo, Italy} %
\author{D.~Vescovi} \affiliation{Goethe University Frankfurt, Germany} \affiliation{Istituto Nazionale di Astrofisica - Osservatorio Astronomico d'Abruzzo, Italy} %
\author{N.~Liu} \affiliation{Institute for Astrophysical Research, Boston University, Boston, USA}
% ...
%
\collaboration{The n\_TOF Collaboration (www.cern.ch/ntof)} \noaffiliation 

\date{\today}

\begin{abstract}
Isotopic measurements of presolar silicon carbide grains from dying stars have revealed a puzzling overabundance of $^{94}$Mo that stellar nucleosynthesis models have failed to reproduce for two decades. This discrepancy challenged our understanding of the slow neutron-capture process ($s$-process) that forges approximately half of the elements heavier than iron. The key uncertainty lies at $^{94}$Nb, a radiactive branching point where competition between neutron capture and beta decay governs the $^{94}$Mo production, yet the neutron-capture cross section had never been measured. Here we report the first experimental determination of the $^{94}$Nb(n,$\gamma$)$^{95}$Nb cross section important for Mo isotopic abundances. The measurement was enabled by a coordinated effort involving high-purity target preparation at Institute of Solid State and Materials Research (IFW) Dresden, radioactive sample production at the Institut Laue-Langevin (ILL) Grenoble, radiochemical characterization at Paul Scherrer Institute (PSI) Villigen, and the Time-of-Flight CERN n\_TOF facility using for the first time segmented total-energy detectors. Incorporation of the resulting Maxwellian-averaged cross section into fully coupled nucleosynthesis models of low-mass asymptotic giant branch (AGB) stars brings them into agreement with the presolar grain data. These results remove a major nuclear-physics input uncertainty at the $^{94}$Nb branching point and provide a firmer foundation for understanding the origin of $^{94}$Mo in the solar system.

\end{abstract}

\maketitle


\textit{Introduction---} Primitive meteorites preserve a record of the  nucleosynthesis processes that built the heavy elements in our Galaxy. Among their most remarkable constituents are presolar grains, microscopic mineral condensates that formed in the outflows of evolved stars and survived incorporation into the Solar System. These grains carry isotopic signatures that provide direct, quantitative tests of stellar nucleosynthesis models, particularly for the slow neutron-capture process ($s$-process) responsible for approximately half of the elements beyond iron.
The majority of SiC grains recovered from primitive meteorites carry distinctive \emph{s}-process isotopic signatures of heavy elements such as Mo, firmly establishing their origin in low-mass ($\sim$1.5--3 $M_{\odot}$) asymptotic giant-branch (AGB) stars, the main stellar site for the \emph{s}-process \cite{Liu24,Speck09,Zinner14}. 
Because Mo is efficiently synthesized by the \emph{s}-process and repeatedly dredged up into the envelope, the surface Mo isotopic composition--- recorded by SiC grains condensing from the surface outflow---directly reflects the interior \emph{s}-process conditions with minimal sensitivity to the initial stellar mass composition. As a result, the Mo isotope ratios preserved in SiC grains provide a direct and sensitive probe of AGB \emph{s}-process nucleosynthesis.

\begin{figure}[!htbp]
    \centering
    \includegraphics[width=1.0\linewidth]{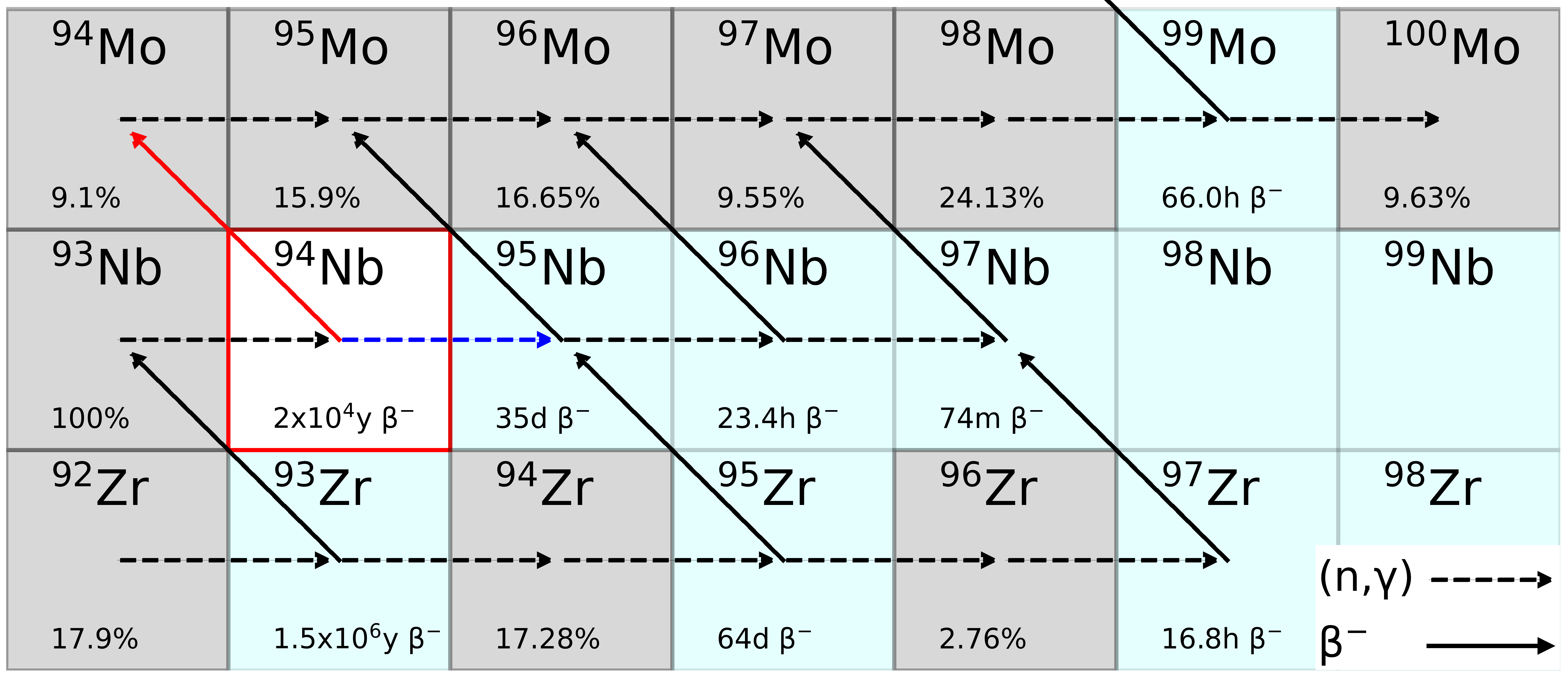}
    \caption{The \emph{s}-process path along the Zr-Nb-Mo region,i.e $^{93}$Zr and $^{94}$Nb. In the figure is highlighted the $^{94}$Nb branching point.}
    \label{fig:sprocess}
\end{figure}

Isotopic analyses of AGB-origin SiC grains over the past two decades consistently indicate that the \emph{s}-process significantly increases the initial $^{94}$Mo abundance with respect to $^{92}$Mo \cite{Lugaro03,Liu19,Stephan19}. However, most post-processing calculations coupled to AGB stellar models predict insufficient productions of $^{94}$Mo by the $s$-process and therefore fall short of explaining the grain data \cite{Lugaro03,Liu19,Palmerini21}. Along the \emph{s}-process path (Fig.~\ref{fig:sprocess}), the production of $^{94}$Mo is sensitive to the branch point at $^{94}$Nb, whose fate depends on the competition between neutron capture and $\beta^{-}$ decay ($T_{1/2}$ = 2$\times$10$^{4}$ years at terrestrial conditions) and thus on the $^{94}$Nb($n$,$\gamma$)$^{95}$Nb reaction rate. At typical \emph{s}-process temperatures ($\sim$3$\times$10$^{8}$ K), the $\beta^{-}$-decay half-life of $^{94}$Nb is predicted to decrease to less than a year \cite{Takahashi87}, while the $^{94}$Nb($n$,$\gamma$)$^{95}$Nb rate remains purely theoretically estimated and unverified experimentally \cite{Dillmann08,Dillmann14}. 
Consequently, accurate measurements of both, the $^{94}$Nb($n$,$\gamma$)$^{95}$Nb cross section and the temperature-dependent $\beta^{-}$-decay rate of $^{94}$Nb, are essential for constraining the \emph{s}-process contribution from this branching point to $^{94}$Mo and improving the reliability of AGB nucleosynthesis models.

In this work, we report the first experimental determination of the $^{94}$Nb(n,$\gamma$)$^{95}$Nb cross section, measured using a radioactive sample of $^{94}$Nb at the high-flux EAR2 station of the CERN n\_TOF facility~\cite{Weiss15,Sabate17,Lerendegui23,Pavon25}. The production and characterization of the $^{94}$Nb sample represented one of the most challenging aspects of this measurement, only accomplished through a collaborative effort involving the ILL high-flux reactor for activation of a high-purity $^{93}$Nb sample, which was produced at IFW-Dresden. Radiochemical characterization of the activated Nb-sample was accomplished at PSI. We present the experimentally-determined Maxwellian-averaged cross sections (MACS) and assess their astrophysical implications by incorporating the new reaction rates into AGB stellar models computed with the FRUITY code\,\cite{Vescovi20}. These models employ a full nuclear network extending up to bismuth, coupled to the physical evolution of the stellar structure, in which the formation of the $^{13}$C pocket ---the primary neutron source in AGB stars--- is attributed to mixing induced by buoyant magnetic flux tubes.

\textit{Experiment---} The synthesis of ultra-high-purity $^{93}$Nb material for the sample used in this experiment was originally conducted at IFW Dresden utilizing refined molten-salt electrolysis, zone melting, and high- and ultra-high-vacuum heat treatments~\cite{Koethe00}. An amount of 304 mg of $^{93}$Nb material with less than 1 ppm of Ta was obtained, consisting of two thin wires, 45 and 47 mm in length and 0.8 mm in diameter. This material was subsequently neutron activated in the V4 irradiation tube at the high-flux ILL reactor in Grenoble for 51 days, corresponding to a thermal neutron fluence of $\approx 4.3 \times 10^{21}$~cm$^{-2}$.

After irradiation the Nb wires were cut into small pieces, distributed in a circle of 16 mm diameter, and sealed from both sides using 25 $\mu$m Kapton foil.
A careful characterization of the activated sample was performed at PSI Villigen using a customized HPGe set-up. The $\gamma$-ray spectroscopy analysis yielded an activity of 10.1 MBq of $^{94}$Nb with no trace of radioactive contaminants. This activity corresponds to 9.24(18)$\times$10$^{18}$ atoms, approximately 1\% of the total number of atoms in the $^{93}$Nb-$^{94}$Nb bulk sample.


For the data analysis (see below) the rather unconventional geometry of the sample could be approximated as multiple thin cylinders irradiated perpendicular to their long axis. Additionally, the high $\beta^{-}$ radioactivity ($Q_{\beta}$= 2045.0 keV followed by a two-step $\gamma$-ray cascade of 702 and 871 keV) and the low amount of $^{94}$Nb introduce important challenges for a time-of-flight ($n$,$\gamma$) measurement. These difficulties were mitigated by using the high instantaneous neutron flux station EAR2 at CERN n\_TOF~\cite{Weiss15,Sabate17,Lerendegui23,Pavon25} in combination of a highly-segmented detection apparatus called sTED. The latter is based on an array of relatively small cells of C$_6$D$_6$, which were designed for rapid response and high count-rate conditions~\cite{Alcayne23,Alcayne24}. This system allows one to handle the high instantaneous count rates at EAR2 during each neutron pulse, as well as the effects of the $\gamma$-ray flash at the short flight-path of EAR2, thus enabling a wide neutron energy range where the detectors can operate with stable and well-controlled performance. Nine sTED detection units were set up in an unconventional compact-ring configuration, which permitted to maximize the signal-to-background ratio~\cite{Balibrea25,Balibrea23}. 

The neutron capture yield was measured by detecting the prompt $\gamma$-rays emitted by the ($n$,$\gamma$) capture reactions. The pulse height weighting technique (PHWT)~\cite{Macklin62} was employed during data processing to ensure that the efficiency of detecting the $\gamma$-ray cascade is proportional to the total energy of the cascade, $E_{C}$, thus rendering it independent of the specific de-excitation path of each individual cascade. After applying the PHWT, the experimental yield is calculated as:

\begin{equation}
Y_{exp}(E_{n}) = f_{N} \frac{C_{W}(E_{n}) - B_{W}(E_{n})}{E_{C} \cdot \phi(E_{n})}
\end{equation}

Here, $C_{W}(E_{n})$ and $B_{W}(E_{n})$ represent the number of counts after application of the PHWT from the sample and background, respectively. Further, $\phi(E_{n})$ denotes the n\_TOF EAR2 neutron fluence~\cite{Sabate17,Lerendegui23,Pavon25}, while $f_{N}$ is the absolute normalization factor for the yield. The latter was determined using the well-established $^{197}$Au($n$,$\gamma$) 4.9 eV saturated resonance method~\cite{Macklin79,Abbondanno04}. For this purpose, high-purity Au and $^{93}$Nb samples with well-defined disk geometries were utilized, having diameters of 16.34 mm and thicknesses of 50 $\mu$m for Au and 2.1 mm for $^{93}$Nb. The methodology experimentally validated at n\_TOF~\cite{Tain02,Abbondanno04} was applied to account for the loss of part of the $\gamma$-ray cascade due to experimental thresholds. To this aim detailed Monte Carlo simulations for $^{197}$Au($n$,$\gamma$) and $^{93}$Nb($n$,$\gamma$) were carried out~\cite{Becvar98,Balibrea25}. In addition, corrections for dead time and pile-up effects were rigorously implemented following the procedure described in Ref.~\cite{Balibrea24}. The absolute normalization of the yield calculated by this method was applied to the $^{94}$Nb sample by scaling the two first $^{93}$Nb resonances present in the measured yield. The uncertainty associated with this normalization, which includes cascade corrections, dead time, and the translation to the final sample yields, amounts to 8\%.

The resonance analysis presented in this work was performed using the Bayesian R-matrix code \textsc{SAMMY}~\cite{Wiarda23}. However, the standard yield calculation in this code had to be replaced by a custom implementation, which incorporates multiple-scattering and self-shielding corrections tailored to the non-standard sample geometry. This implementation follows the methodology described in Ref.~\cite{Mancinelli12} and is essential for determining accurate $^{94}$Nb resonance parameters. Additional details on the analysis and the results for the $^{94}$Nb and $^{93}$Nb samples measured in this work will be presented in a forthcoming publication~\cite{Balibrea26}.

In total, we have observed and analyzed ten 
$^{94}$Nb($n$,$\gamma$) resonances up to a neutron energy of about 800~eV. For the first resonance we were able to determine a radiation width of  $\Gamma_{\gamma}$=145(40) meV. Where necessary, this value was fixed for the other resonances during the determination of other parameters. A summary of the resonance parameters determined in this work is available in the supplementary material.

Taking into account the observed resonances and the detection threshold, the fraction of missing resonances was estimated, yielding an $s$-wave resonance spacing of $D_{0}$=32(7) eV and a neutron strength function of $S_{0}$=0.19(8)10$^{-4}$. The MACS at different stellar temperatures were calculated using the average resonance parameters, following the approach of Refs.~\cite{Guerrero20,Casanovas24}. In practice, $p$-wave neutrons contribute significantly to the MACS of $^{94}$Nb($n$,$\gamma$)$^{95}$Nb. This contribution was evaluated by assuming the same average radiative width, $\Gamma_{\gamma}$, as for $s$-wave neutrons—an approximation fully supported by statistical-model calculations \cite{Becvar98,Koning23}—and by adopting a neutron strength function $S_{1}$ based on values from neighboring isotopes. $D_{1}$ was estimated under the assumption of parity-independent level densities, using the spin cut-off parameter described in Ref.~\cite{Egidy09}.

\begin{figure}
    \centering
    \includegraphics[width=1.0\linewidth]{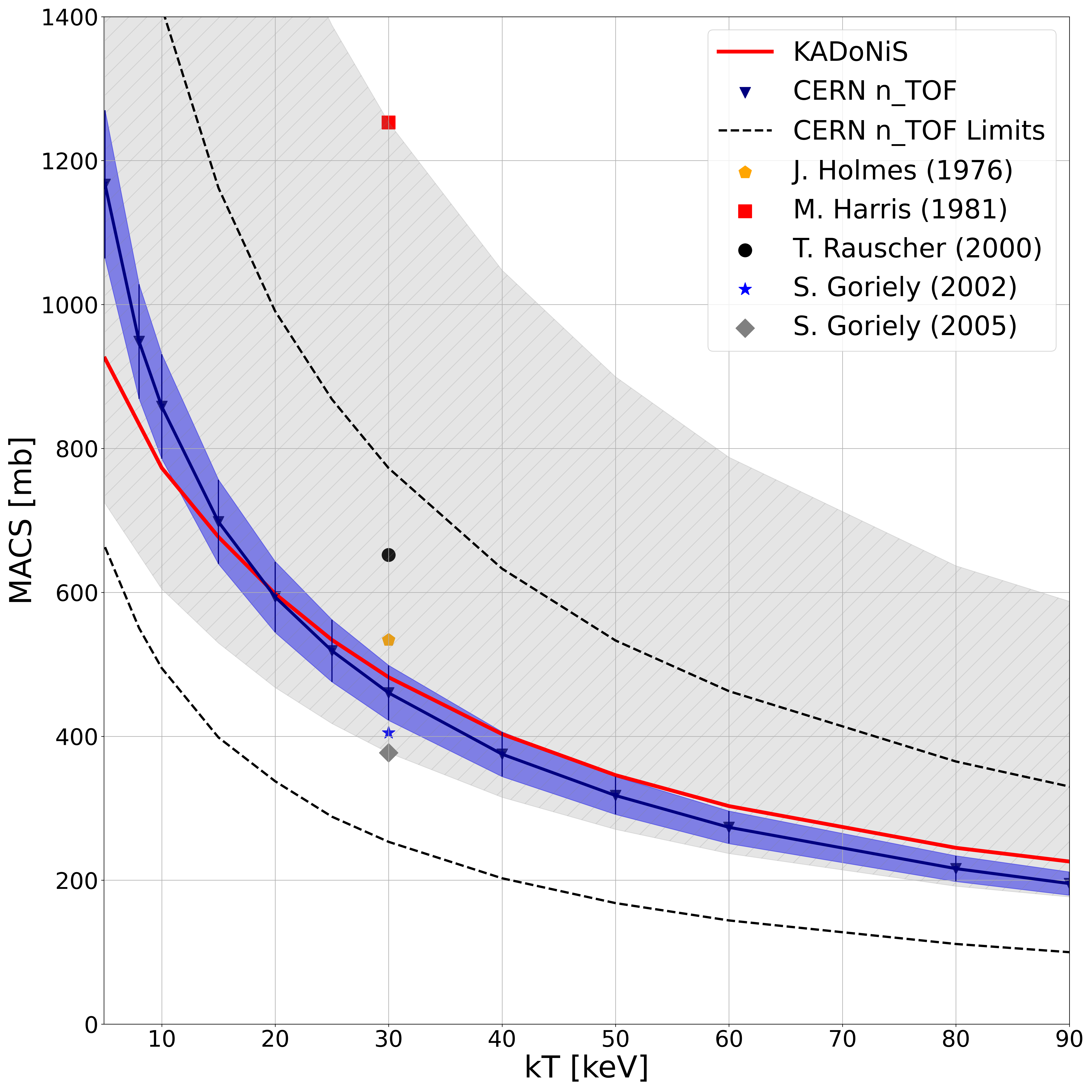}
    \caption{MACS as a function of $kT$ obtained from experimental data, compared with purely KADoNiS estimation. See text for details.}
    \label{fig:macs}
\end{figure}

\textit{Results---} Fig.~\ref{fig:macs} compares the MACS as a function of stellar thermal energy ($kT$) obtained in this work with the values adopted by KaDoNis v0.3~\cite{Dillmann08,Dillmann14}. Because the latter are purely theoretical, no uncertainty is displayed for them except at 30 keV. Uncertainties presented from n\_TOF -as a blue band- correspond to the systematic uncertainty of 8\%, arising from the different experimental corrections discussed above. The dashed lines then indicate the $\pm2\sigma$ exclusion limits from the n\_TOF results as inferred from the uncertainties in the average resonance parameters used in the calculations. A more detailed description of this procedure can be found in the Supplementary Material, which also provides the numerical values of the calculated MACS.

The new determined MACS values at $kT$=5 and 30 keV are 1167(93) and 460(37) mb, respectively. They correspond to a 26\% higher value at 5 keV, while at 30 keV the experimental result is within 5\% of the KADoNiS value. At higher $kT$ values, the new MACS are slightly lower than those of KADoNiS, by up to 13\%.



\begin{figure}[!htbp]
    \centering
    \includegraphics[width=1.0\linewidth]{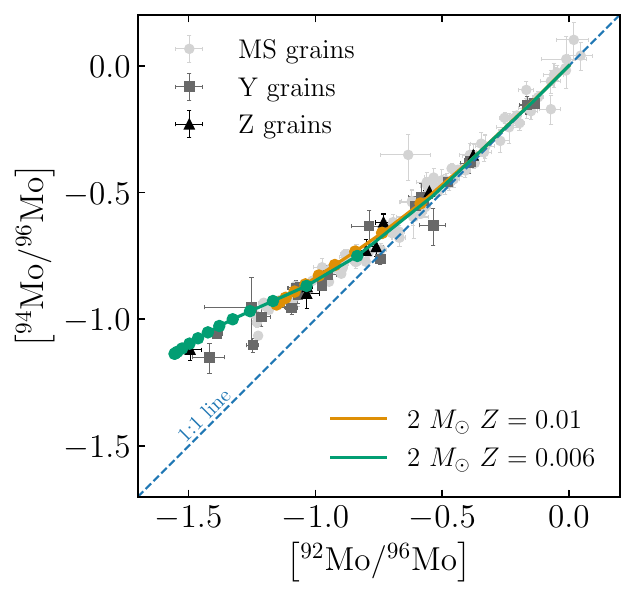}
    \caption{Molybdenum three-isotope plot comparing MS, Y, and Z grains with the $2~M_{\odot}$ magnetic fruity models predictions utilizing the MACS measured in this work. The grain data are ploted with 1 $\sigma$ uncertainties\cite{Liu19, Stephan19,Stephan25}. The lines with symbols trace the trajectory of envelope isotopic compositions at two different metallicities ($Z$), with symbols indicating phases where the envelope becomes C-rich ($C/O>$1), a necessary condition for SiC dust condensation. The $Z_{\odot}$ adopted in the models is 0.0185.}
    \label{fig:grains}
\end{figure}

\textit{Discussion---}The new $^{94}$Nb($n$,$\gamma$)$^{95}$Nb MACS reported here enables a more reliable constraint on one of the key branch points along the Zr--Nb--Mo \emph{s}-process path, when compared to the pure theoretical calculations available thus far. Over the astrophysically relevant range of thermal energies, our MACS values are comparable to or slightly higher than previous theoretical estimates \cite{Dillmann08}, ruling out a significantly slower $^{94}$Nb($n$,$\gamma$)$^{95}$Nb rate as a mechanism to enhance the \emph{s}-process production of $^{94}$Mo in low-mass AGB stars. 
Grains with inferred AGB origins comprise the mainstream (MS) ($\sim$90\%), Y ($\sim$1--5\%), and Z ($\sim$1--5\%) groups, differing in their Si isotopic compositions, but exhibiting indistinguishable Mo isotopic ratios \cite{Liu19}. These shared \emph{s}-process Mo patterns show that all these three groups formed in near-solar-metallicity AGB stars\,\cite{Cristallo20}. When incorporated into fully coupled magnetic FRUITY AGB models \cite{Vescovi20}, the overproduction of $^{94}$Mo relative to the initial abundance remains at the level inferred from MS, Y, and Z SiC grains \cite{Liu19,Stephan19,Stephan25}, and the revised model predictions overlap well with the grain data shown in Fig.~\ref{fig:grains}.
During interpulse periods---when the $^{13}$C($\alpha$,$n$)$^{16}$O source provides a low neutron density ($\sim$10$^{7}$--10$^{8}$ cm$^{-3}$) \cite{Gallino98} and the $^{93}$Zr and $^{94}$Nb branch points remain inactive---the $s$-process flow converts the initial amount of $^{93}$Nb directly to $^{94}$Mo, but it is then almost completely destroyed through neutron captures. Conversely, the long-lived $^{93}$Zr is largely produced by neutron captures on $^{92}$Zr.
At the onset of the following thermal pulse (TP), partial activation of the $^{22}$Ne($\alpha$, $n$)$^{25}$Mg neutron source generates a brief episode of high neutron densities (10$^{9}$--10$^{11}$ cm$^{-3}$) \cite{Gallino98}, thereby activating the branch point at $^{94}$Nb and causing neutron capture and $\beta^{-}$ decay to compete.
\begin{figure}[!htbp]
    \centering
    \includegraphics[width=1.0\linewidth]{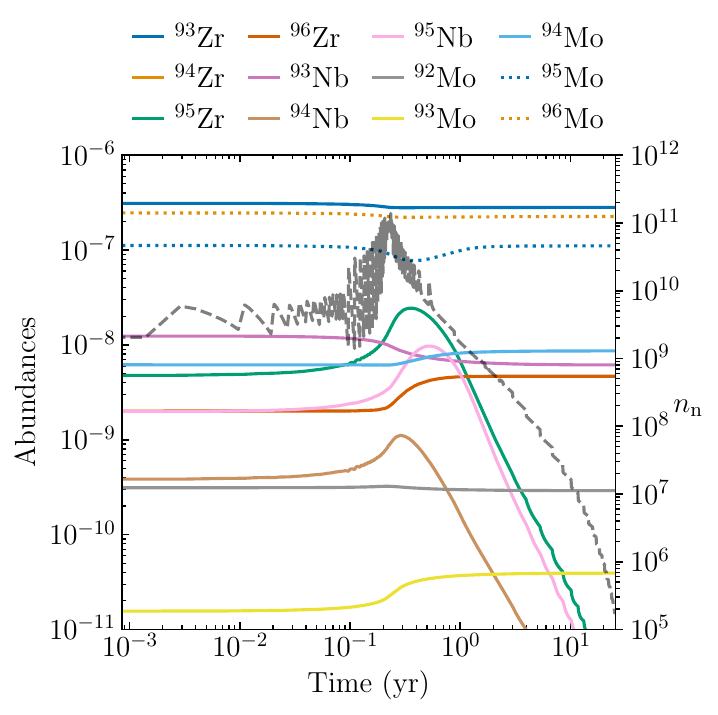}
    \caption{Time-evolution of the key Zr, Nb, and Mo isotopes 
    and the neutron density (dashed gray line) during a representative thermal pulse for the $2~M_{\odot}$ obtained from magnetic fruity models.}
    \label{fig:tp}
\end{figure}
Under these conditions, the abundance of $^{94}$Mo increases during the TP~\cite{Lugaro03}: the large amount of $^{93}$Nb produced by the decay of $^{93}$Zr during the previous interpulse phase is converted into $^{94}$Nb, a significant fraction of which undergoes $\beta^{-}$ decay to $^{94}$Mo (see Figure~\ref{fig:tp}).
The alternative production channel through successive neutron captures on $^{92}$Mo and $^{93}$Mo remains negligible, given the low initial $^{92}$Mo abundance.
At the same time, most of the $s$-process flow from $^{93}$Zr goes to $^{94}$Zr rather than $\beta$-decaying to $^{93}$Nb. The newly formed $^{94}$Zr continues to capture neutrons to form $^{95}$Zr, which subsequently decays to $^{95}$Nb and then to $^{95}$Mo, thus bypassing $^{94}$Mo entirely (Fig.~\ref{fig:sprocess}).
As a result, the $^{94}$Nb($n$,$\gamma$)$^{95}$Nb rate determined in this work confirms that the synthesis of $^{94}$Mo in AGB stars is tightly linked to the short high–neutron–density episodes that occur during thermal pulses. During the long interpulse phases, its production remains inefficient and $^{94}$Mo is largely destroyed.


The FRUITY models also allow assessing the sensitivity of the $^{94}$Mo yield to the $^{94}$Nb branching in a fully coupled stellar framework. Across the range of low-mass, near-solar-metallicity AGB stars relevant for MS, Y, and Z grains, adopting our experimentally determined $^{94}$Nb($n$,$\gamma$)$^{95}$Nb rate in place of the previous theoretical estimate \cite{Dillmann08,Dillmann14} leads to only modest changes in the final $^{94}$Mo/$^{96}$Mo ratios. This limited response indicates that earlier theoretical rates were already close to the MACS determined experimentally. Importantly, with the new rate the FRUITY models reproduce the \emph{s}-process contribution to the $^{94}$Mo abundance inferred from SiC grains (Fig.~\ref{fig:grains}), consistent with the conclusions of a recent postprocessing study \cite{Szanyi25}. Given that the nuclear inputs in most of previous postprocessing calculations \cite{Liu19,Palmerini21} are similar to those adopted in this work, we infer that the longstanding $^{94}$Mo data-model discrepancy likely arose from limitations in their postprocessing treatments---particularly their approximate handling of the temperature- and density-dependent $\beta^{-}$-decay rate of $^{94}$Nb---rather than from unrealistic neutron-capture rates. As the stellar $\beta^{-}$-decay rate remains based solely on theoretical calculations, experimental constraints, e.g., the goal of PANDORA project \cite{Mascali23}, will be crucial for further validating the $^{94}$Mo production pathway in AGB stars.
By providing the first experimental determination of the $^{94}$Nb($n$,$\gamma$)$^{95}$Nb rate, this work removes a significant source of uncertainty in the nuclear input and sets a firmer foundation for future efforts to disentangle the stellar and galactic processes that shaped the $^{94}$Mo inventory in the Solar System. For completeness, measurements on the stable Mo isotopes have recently been conducted at n\_TOF \cite{Mucciola22,Mucciola23} to confirm or update the remaining cross sections involved in the \emph{s}-process production of $^{94}$Mo network with constrained uncertainties.

\textit{Conclusions---} We have performed the first experimental determination of the (n,$\gamma$) cross section on the radioactive isotope $^{94}$Nb, a key branching point in the $s$-process flow toward $^{94}$Mo. This challenging $^{94}$Nb experiment required a comprehensive multi-facility collaboration: production of ultra-pure $^{93}$Nb material at IFW, activation at the ILL high-flux  reactor, radiochemical characterization at PSI, and high-resolution time-of-flight measurements at the CERN n\_TOF EAR2 facility. The use of the newly-developed sTED detection system was essential to achieve the required detection sensitivity. The resulting MACS values represent the first experimental constraints on this critical reaction, with values comparable to, or slightly higher than, previous theoretical estimates.

To assess the astrophysical implications of our results, we incorporated the measured rates into fully-coupled AGB stellar model calculations based on the magnetic-FRUITY framework. The calculations demonstrate that fully coupled modern AGB models employing magnetic buoyancy-induced mixing, when combined with our $^{94}$Nb($n$,$\gamma$)$^{95}$Nb reaction rates, successfully reproduce the $^{94}$Mo/$^{96}$Mo ratios observed in AGB-origin presolar SiC grains. 
Our results confirm that the previously problematic $^{94}$Mo abundances can be solved through the combination of accurate treatment of thermally sensitive $s$-process branchings and experimentally-constrained reaction rates, without requiring exotic astrophysical scenarios.


While our measurement rules out uncertainties in the $^{94}$Nb($n,\gamma$)$^{95}$Nb reaction as a source of model-data discrepancies, the complete picture of $^{94}$Mo production in AGB stars still requires better constraints on the temperature-dependent $^{94}$Nb $\beta$-decay rate under stellar conditions. 
Combined with continued refinements in AGB stellar modeling, these efforts will provide a comprehensive understanding of $s$-process branching-point nucleosynthesis and its signatures in presolar stardust.

\textit{Acknoledgments---} The authors acknowledge support from all the funding agencies of participating institutions. Part of this work was supported by the European Research Council (ERC) under the European Union's Horizon 2020 research and innovation programme (ERC-COG Nr.~681740 and ERC-STG Nr.677497), European H2020-847552 (SANDA), the MCIN/AEI 10.13039/\-501100011033 under grants Severo Ochoa CEX\-2023-001\-292\--S, PID2022-138297NB-C21, PID2019-104714GB-C21, FPA\-2017-83946-C2-1-P, FIS2015-71688-ERC, FPA\-2016-77689-C2-1-R, RTI2018\--098117\--B\--C21, PGC2018\--096717\--B\--C21, PID2021\--123100NB\--I00 funded by MCIN/AEI 10.13039/\-501100011033/\-FEDER, UE, PCI2022\--135037\--2 funded by MCIN/\-AEI 10.13039/\-501100011033 and EU-\-NextGenerationEU/\-PRTR. N.L. acknowledges support from NASA through grants 80NSSC23K1034 and 80NSSC24K0132. S.C. and D.V. acknowledge funding by the European Union – NextGenerationEU RFF M4C2 1.1 PRIN 2022 project "2022RJLWHN URKA" and by INAF 2023 Theory Grant ObFu 1.05.23.06.06 “Understanding R-process and Kilonovae Aspects". P. Z. acknowledges support from Croatian Science Foundation project IP-2022-10-3878.

The corresponding author has applied a Creative Commons Attribution (CC BY-NC-ND 4.0) license.

\bibliographystyle{apsrev4-2}
\bibliography{bibliography}


\end{document}